\pdfoutput=1
\documentclass[12pt,draftcls,onecolumn]{IEEEtran}
\usepackage{flushend}
\usepackage[super,comma, sort&compress]{natbib}
\usepackage{array}
\usepackage{amsmath,amssymb}    
\usepackage[dvips]{graphicx}   
\usepackage{verbatim}   
\usepackage{color}      
\usepackage{subfigure}  
\usepackage{booktabs}
\usepackage{multirow}
\usepackage{graphics}
\usepackage{color}
\usepackage{etoolbox}
\makeatletter
\patchcmd{\@makecaption}
  {\scshape}
  {}
  {}
  {}
\makeatother
\usepackage{epsfig}
\usepackage{float}
\usepackage{algorithm}
\usepackage{algorithmic}
\usepackage{subfigure}
\makeatletter
\newcommand*{\rom}[1]{\expandafter\@slowromancap\romannumeral #1@}

\makeatother
\IEEEoverridecommandlockouts
\begin{document}
\title{\textbf{Reliable and Explainable Machine Learning Methods for Accelerated Material Discovery}}
\author{\IEEEauthorblockN{Bhavya Kailkhura*, Brian Gallagher, Sookyung Kim, Anna Hiszpanski, T. Yong-Jin Han*\\}
\IEEEauthorblockA{Lawrence Livermore National Laboratory\\}
\textnormal{Corresponding Authors: B.K. (kailkhura1@llnl.gov); T.Y.H. (han5@llnl.gov)}
}

\maketitle
\begin{abstract}
\end{abstract}
Material scientists are increasingly adopting the use of machine learning (ML) for making potentially important decisions, such as, discovery, development, optimization, synthesis and characterization of materials.
However, despite ML's impressive performance in commercial applications, several unique challenges exist when applying ML in materials science applications. In such a context, the contributions of this work are twofold. 
First, we identify common pitfalls of existing ML techniques when learning from underrepresented/imbalanced material data. Specifically, we show that with imbalanced data, standard methods for assessing quality of ML models break down and lead to misleading conclusions. Furthermore, we find that the model's own confidence score cannot be trusted and model introspection methods (using simpler models) do not help as they result in loss of predictive performance (reliability-explainability trade-off).
Second, to overcome these challenges, we propose a general-purpose explainable and reliable machine-learning framework. Specifically, we propose a novel pipeline that employs an ensemble of simpler models to reliably predict material properties. We also propose a transfer learning technique and show that 
the performance loss due to models' simplicity can be overcome by exploiting correlations among different material properties. A new evaluation metric and a trust score to better quantify the confidence in the predictions are also proposed. 
To improve the interpretability, we add a rationale generator component to our framework which provides both model-level and decision-level explanations.
Finally, we demonstrate the versatility of our technique on two applications: $1)$ predicting properties of crystalline compounds, and $2)$ identifying novel potentially stable solar cell materials. We also point to some outstanding issues yet to be resolved for a successful application of ML in material science.


\section{Introduction}
\label{intro}
\subsection{Motivation}
Driven by the success of machine learning (ML) in commercial applications (e.g., product recommendations and advertising), there are significant efforts to exploit these tools to analyze scientific data. One such effort is the emerging discipline of Materials Informatics which applies ML methods to accelerate the selection, development, and discovery of materials by learning structure-property relationships. Materials Informatics researchers are increasingly adopting ML methods in their workflow to predict materials' physical, mechanical, optoelectronic, and thermal properties (e.g., crystal structure, melting temperature, formation enthalpy, band gap).
While commercial use cases and material science applications may appear similar in their overall goals, we argue that fundamental differences exist in the corresponding data, tasks, and requirements. Applying ML techniques without careful consideration of their assumptions and limitations may lead to missed opportunities at best and a waste of substantial resources and incorrect scientific inferences at worst. In the following, we mention unique challenges that the Materials Informatics community must overcome for universal acceptance of ML solutions in material science.

\begin{figure}[t] 
  \centering
    \includegraphics[width=1.0\textwidth]{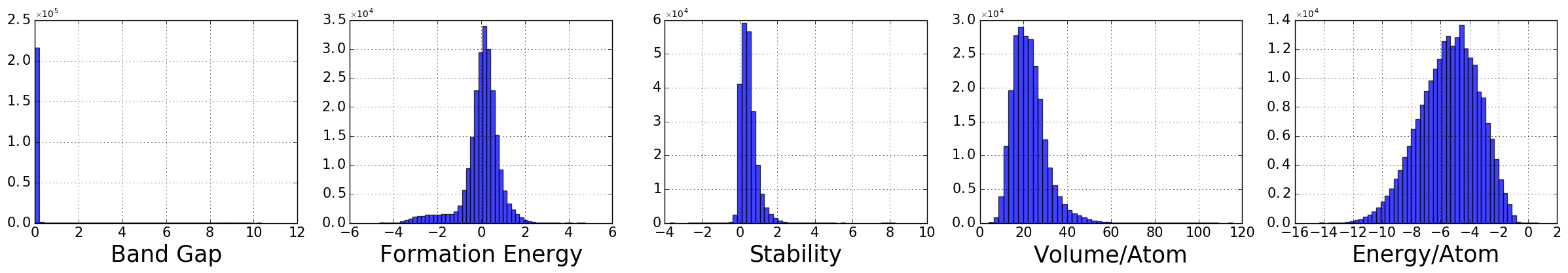}
    \caption{Histograms (number of compounds vs. targeted property bin) of targeted properties of the OQMD database show heavily skewed distributions. {We show that conventional machine learning approaches: (a) produce inaccurate inferences in sparse regions of the property-space and (b) are overconfident in the accuracy of such predictions. The proposed approach overcomes these shortcomings.}}
        \label{imb_hist}
\end{figure}
\vspace{0.1in}
\noindent\textbf{Learning From Underrepresented and Distributionally Skewed Data:}
One of the fundamental assumptions of current ML methods is the availability of densely and uniformly sampled (or balanced) training data. When there is an under-representation of certain classes in the data, standard ML algorithms provide incorrect inferences across the classes of the data.
Unfortunately, in most material science applications, balanced data is exceedingly rare, and virtually all problems of interest involve various forms of extrapolation due to underrepresented data and severe class distribution skews.
As an example, materials scientists are often interested in designing (or discovering) compounds with uncommon targeted properties, e.g., high $T_C$ superconductivity or large $ZT$ for improved thermoelectric power~\cite{theory}, shape memory alloys (SMAs) with the targeted property of very low thermal hysteresis~\cite{accelerated}, and band gap energy in the desired range ($0.9-1.7$ eV) for solar cells~\cite{ward2016general}. In such applications, we encounter highly imbalanced data (with targeted materials being in the minority class) due to these design choices or constraints. 
Consider a task of predicting material properties (e.g., bandgap energy, formation energy, stability, etc.) from a set of feature vectors (or descriptors) corresponding to crystalline compounds.   
One representative database for such a data set is the Open Quantum Materials Database (OQMD)\cite{oqmd}, which contains several properties of crystalline compounds as calculated using density functional theory (DFT). Note that, the OQMD database contains data sets with strongly imbalanced distributions of target variables, i.e., material properties. 
In Figure~\ref{imb_hist}, we plot the histogram of several commonly targeted properties. It can be seen that, the data set exhibits severe distribution skews. For example, $95\%$ of the compounds in the OQMD are possibly conductors with band gap value equal to zero. 
Note that if the sole aim of the ML model is to maximize overall accuracy, the ML algorithm will perform quite well by ignoring or discarding the minority class. 
However, in practice, correctly classifying and learning from the minority class of interest may be more important than possibly misclassifying the majority classes.  

\noindent\textbf{Explainable ML Methods without Compromising the Model Accuracy:}
A common misconception is that increasing model complexity can address the challenges of underrepresented and distributionally skewed data. However, this can only superficially solve some of these problems. Furthermore, increasing the complexity of ML models may increase the overall accuracy of the system at the cost of making the model very hard to interpret. Understanding why an ML model made a certain prediction or recommendation is crucial, since it is this understanding that provides the confidence to make a decision and that will lead to new hypotheses and ultimately new scientific insights. 
%
%
%
Most of the existing approaches define explainability as the inverse of complexity and achieve explainability at the cost of accuracy. This introduces a risk of producing explainable but misleading predictions. With the advent of highly predictive but opaque ML models, it has become more important than ever to understand and explain the predictions of such models and to devise explainable scientific machine learning techniques without sacrificing predictive power.  

\noindent\textbf{Better Evaluation and Uncertainty Quantification Techniques for Building Trust in ML:}
For a credible use of ML in material science applications, we need the ability to rigorously quantify the ML performance. Traditionally, the quality of an ML model is measured by the accuracy on test data using cross-validation. Considering the scarcity of densely sampled data in most material science problems, high accuracy on the test data can hardly provide confidence on the quality and generality of ML systems. A natural solution is to use a model's own reported confidence (or uncertainty) score for quantifying trust in the prediction. However, a model's confidence score alone may not be very reliable.
For example, in computer vision, well-crafted perturbations to images can cause classifiers to make mistakes (such as, identifying a panda as a gibbon or confusing a cat with a computer) with very high confidence\cite{bhavya_uni}. As we will show later, this problem also persists in the Materials Informatics pipeline (especially with distributional skewness). Nevertheless, knowing when a classifier's (or regressor's) prediction can be trusted is useful in several other applications for building assured ML solutions. Therefore, we need to augment current validation techniques with additional components to quantify generalization performance of scientific ML algorithms and devise reliable uncertainty quantification methods to establish trust in these predictive models.

\subsection{Literature Survey}
In the recent past, the materials science community has used ML methods for building predictive models for several applications\cite{ms1,ms2,ms3,ms4,ms5,ms6,ms7,ms8,ms9,ms10,ms11,ms12,ms13,ms14}. 
Seko et al.\cite{ms9} considered the problem of building ML models to predict the melting temperatures of binary inorganic compounds. The problem of predicting the formation enthalpy of crystalline compounds using ML models was considered recently\cite{ms2,ms3,ap1}. Predictive modeling for crystal structure formation at a certain composition are also being developed\cite{ms4,ap2,ap3,ap4}. The problem of band gap energy prediction of certain classes of crystals\cite{ap5,ap6} and mechanical property prediction of metal alloys was also considered in the literature\cite{ms12,ms13}. Ward et al.\cite{ward2016general} proposed a general-purpose ML framework to predict diverse properties of crystalline and amorphous materials, such as band gap energy and glass-forming ability.

Thus far, the research on applying ML methods for material science applications has predominantly focused on improving overall accuracy of predictive modeling. However, imbalanced learning, explainability and reliability of ML methods in material science have not received any significant attention. As mentioned earlier, these aspects pose a real problem in deriving correct and reliable scientific inferences and the universal acceptance of machine learning solutions in material science, and deserves to be tackled head on.
    
\subsection{Our Contributions}
In this paper, we take some first steps in addressing the challenge of building reliable and explainable ML solutions for Materials Informatics applications. The main contributions of the paper are twofold. 
First, we identify some shortcoming with training, testing, and uncertainty quantification steps in existing ML techniques while learning from underrepresented
and distributionally skewed data. Our finding raises serious concerns regarding the reliability of existing Materials Informatics pipelines. 
Second, to overcome these challenges, we propose a general-purpose explainable and reliable machine-learning methods for enabling reliable learning from underrepresented and distributionally skewed data. We propose the following solutions: $1)$ novel learning architecture to bias the training process to the goals of imbalanced domains; and $2)$ sampling approaches to manipulate the training data distribution so as to allow the use of standard ML models; $3)$ reliable evaluation metrics and uncertainty quantification methods to better capture the application bias. 
To improve the explainability, as oppose to other existing approaches which train an independent regression model per property, we employ a simple and computationally cheap partitioning scheme. This scheme first partitions the data into sub classes of materials based on their property values and train separate simpler regression models for each  group. Note that our approach differs in its motivation (and operation) from similar concept utilized by Ward et al.\cite{ward2016general}. Our motivation behind partitioning is to enhance the ``explainability", as opposed to the previous approach \cite{ward2016general}, where a computationally expensive exhaustive search was performed to find artificial groups to enhance the accuracy of predictions.  
In our case, our explainability enhancing partitioning scheme in fact hurts our predictive performance (or accuracy). To compensate this performance loss, we utilize transfer learning by exploiting correlation among different material properties to improve the regression performance. We show that the proposed transfer learning technique can overcome the performance loss due to simplicity of the models.
To further improve the interpretability of the ML system, we add a rationale generator component to our framework. The goal of the rationale generator is twofold: $1)$ provide explanations corresponding to an individual prediction, and $2)$ provide explanations corresponding to the regression model. For individual prediction, the rationale generator provides explanations in terms of prototypes (or similar but known compounds). This helps a material scientist to use his/her domain knowledge to verify if similar known compounds or prototypes satisfy the requirements or constraints imposed. On the other hand, for regression models, the rationale generator provides global explanations regarding the whole material sub-classes. This is achieved by providing feature importance for every material sub-class. Finally, we propose a new evaluation metric and a trust score to better quantify confidence and establish trust in the ML predictions.
We demonstrate the applicability of our technique by using it for two applications: $1)$ predicting five physically distinct properties of crystalline compounds, and $2)$ identifying potentially stable solar cells. 

\section{Results and Discussions}
First, we discuss proposed ML method with a focus on reliability and explainability using the data from the Open Quantum Materials Database (OQMD). Next, we demonstrate the application of our approach in two material science problems.

\subsection{General-Purpose Reliable and Explainable ML Framework}

\begin{figure}[t] 
  \centering
    \includegraphics[width=1.0\textwidth]{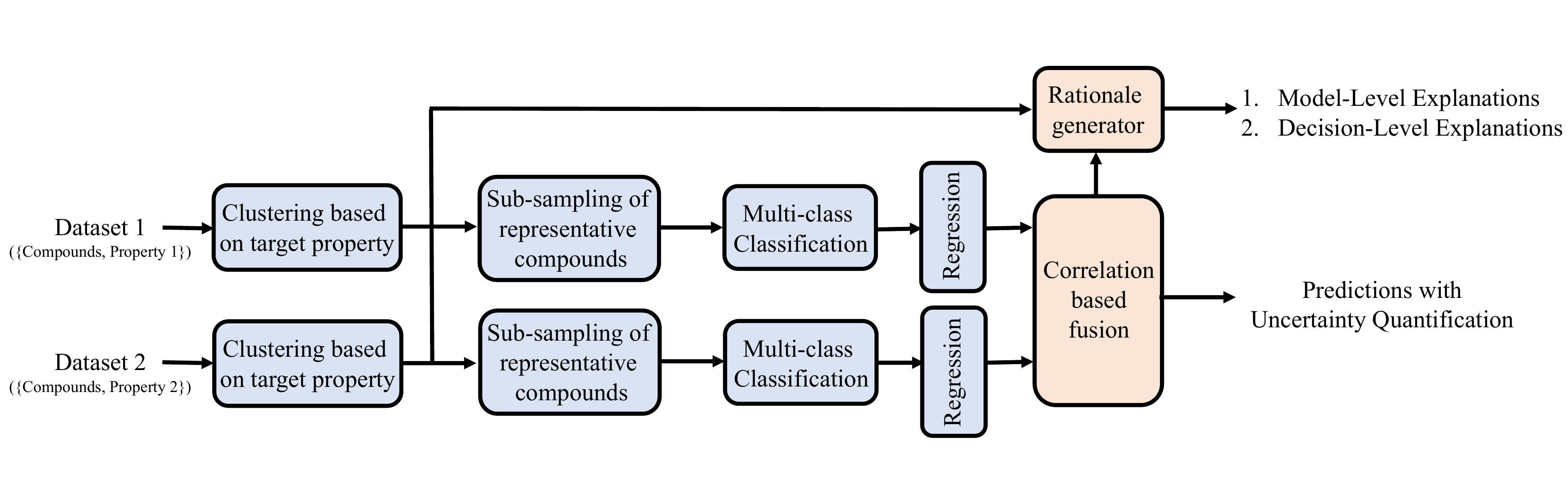}
    \caption{An illustration of proposed ML pipeline for material property prediction.}
    \label{ml_f}
\end{figure}

To solve the problem of reliable learning and inference from distributionally skewed data, we propose a general purpose ML framework (see Fig.~\ref{ml_f}).
Instead of developing yet another ML algorithm to improve accuracy for a specific application, our objective is to develop generic methods to improve reliability, explainablity and accuracy in the presence of imbalanced data.
The proposed framework is agnostic to the type of training data, can utilize a variety of already-developed ML algorithms, and can be reused for a broad variety of material science problems. The framework is composed of three main components: $1)$ novel training procedure for learning from imbalanced data, $2)$ rationale generator for model-level and decision-level explainability, and $3)$ reliable testing and uncertainty quantification techniques to evaluate the prediction performance of ML pipelines.

\subsubsection{Training Procedure}
\label{train_app}
Building an ML model for materials properties prediction can be posed as a regression problem where the goal is to predict  continuous valued property values from a set of material attributes/features. The challenge in our target task is that due to the presence of distributional skewness,  ML models do not generalize well (specifically in domains which are not well represented using available labeled data (or minority classes)). To solve this problem, we propose a generic ML training process that is applicable to a broad range of materials science applications which suffer from the distributionally skewed data.
We will explain the proposed training process with the help of following running example: A material scientist is interested in learning an ML model targeting a specific class of material properties, e.g., stable wide bandgap materials in a certain targeted range. 
In most of the cases, we have domain knowledge about the range of property values for specific classes of materials, e.g., conductors have bandgap energies equal to zero, typical semiconductors have bandgap energies in the range of $1.0$ to $1.5$ eV, whereas wide bandgap materials have bandgap energies greater than $2$ eV. 
These requirements introduce a partition of the property space in multiple material classes\footnote{This partition can also be introduced artificially by imposing constraints on the gradient of the property values so that compounds with similar property value are in the same class.}.  
Given $N$ training data samples $\{X_i,(Y_i^1,\cdots,Y_i^M)\}_{i=1}^N$, where, $X_i$ is feature/attribute vector and $Y_i^j$ is $j^\text{th}$ property value corresponding to compound $i$,
the steps in the proposed training procedure are as follows:

\begin{enumerate}
\item Partition the property space in $K$ regions/classes and obtain transformed training data samples $\{X_i,(Z_i^1,\cdots, Z_i^M)\}_{i=1}^N$ where $Z_i^j \in \{1,\cdots, K\}$.
\item For each property in $j \in \{1,\cdots,M\}$, perform sub-sampling\footnote{Other sophisticated sampling techniques\cite{smote} or generative modeling approaches can also be used.} on sample compounds in $K$ distinct classes, and obtain an evenly distributed training set: $\{X_i,Z_i^j\}_{i=1}^{N_j}$. 
\item Train $M$ multi-class classifiers (one per property) on balanced datasets $\{X_i,Z_i^j\}_{i=1}^{N_j}$ to predict which class a compound belongs to.
\item For every $(j,k)$ pair, train a regressor on $\{X_i,Y_i^j\}_{i=1}^{N_j}$ to predict property values $\hat{Y}_i^j$.
\item Finally, utilize correlation among properties to improve the model accuracy by employing transfer learning (explained next).
\end{enumerate}

At the test time, to predict $j^\text{th}$ property of the test compound, the ML algorithm first identifies the class the test compound belongs to by using trained $j^\text{th}$ multi-class classifier. Next, depending on the predicted class $k$ for property $j$, $(j,k)^\text{th}$ regressor is used, along with transfer learning step, to predict property values of the test compound. Next, we provide details and justifications for each of these steps in our ML pipeline.    

\noindent Steps $1$ to $3$ transform a regression problem into a multi-class classification problem on sub-sampled training data. 
The change that is carried out has the goal of balancing the distribution of the least represented (but more important) material classes with the more frequent observations\footnote{Note that the proposed framework is general enough to utilize other sophisticated imbalanced learning strategies (such as, ensemble learning, data pre-processing and cost-based learning) to further improve the performance.}. Furthermore, instead of having a single model trained on the entire training set, having smaller and simpler models for different classes of materials helps to gain better understanding of sub-domains using the rationale generator (explained later).

\noindent Next, we explain the proposed transfer learning technique which exploits correlations presented among different material properties to improve the regression performance. We devise a simple knowledge transfer scheme to utilize the marginal estimates/predictions from step $4$ where regressors were trained independently for different properties. Note that, for each compound $i$, we get an independent estimate $\mathbf{\hat{Y}_i}\approx\{\hat{Y}_i^1,\cdots,\hat{Y}_i^M\}$ from step $4$. In step $5$, we augment the original attribute vector $X_i$ with independent estimates $\mathbf{\hat{Y}_i}$ and use it as a modified attribute vector and train regressors for each $(j,k)$ pair. We found that this simple knowledge transfer scheme significantly improves the regression performance. 

\subsubsection{Rationale Generator}
The goal of rationale generator is to provide: $(a)$ decision level explanations, and $(b)$ model level explanations. Decision level explanations provide reasoning such as: what made an ML algorithm make a specific decision/prediction? 
On the other hand, model level explanations are focused on providing understandings at the class level, e.g., which chemical attributes help in discriminating among insulators, semi-conductors, and conductors?

\noindent\textit{Decision Level Explanations:} The proposed ML pipeline explains its predictions for previously unseen compounds by providing similar known examples (or prototypes).
Explanation by examples is motivated by the observation that studies of human reasoning have shown that the use of examples (analogy) is fundamental to the development of effective strategies for better decision-making\cite{human_proto}. Example-based explanations are widely used in the effort to improve user explainability of complex ML models. In our context, for every unseen test example, in addition to predicted property values, we provide similar experimentally known compounds with corresponding similarity to the test compound in the feature space. Our feature space is heterogeneous (both continuous and categorical features), thus, Euclidean distance is not reliable. Thus, we propose to quantify similarity using Gower's metric\cite{gower}. Gower's metric can be used to measure similarity between data containing a combination of logical, numerical, categorical or text entries. The distance is always a number between $0$ (similar) and $1$ (maximally dissimilar).
Furthermore, as a consequence of breaking a large regression problem into a multi-class classification followed by a simpler regression problem, we can also provide a logical sequence of decisions taken to reach a prediction. 

\noindent\textit{Model Level Explanations:} 
Knowing which chemical attributes are important in a model's prediction (feature importance) and how they are combined can be very powerful in helping material scientists understand and trust automatic ML systems.
Due to the structure of our pipeline (regression+classification), we can provide a more fine grained feature importance explanations compared to having a single regression model.
Specifically, we break the feature importance of attributes to predict a material property into: $1)$ feature importance for discriminating among different material classes (inter-class),  and $2)$ feature importance for regression on a material sub-domain (intra-class). This provides a more in depth explanation of the property prediction process. 

\subsubsection{Robust Model Performance Evaluation and Uncertainty Quantification}

The distributionally skewed training data biases the learning system towards solutions that may not be in accordance with the user's end goal.
Most existing learning systems work by searching the space of possible models with the goal of optimizing some criteria (or numerical score). These metrics are usually related to some form of average performance over the whole train/test data and can be misleading in cases where sampled train/test data is not representative of the true distribution. More specifically, commonly used evaluation metrics (such as mean squared error, R-squared error, etc.) assume an unbiased (or uniform) sampling of the test data and break down in the presence of distributionally skewed test data (shown later). 
Therefore, we propose to perform class specific evaluations (by partitioning the property space into multiple classes of interest) which better characterizes the predictive performance of ML models in the presence of distributionally skewed data. We also recommend visualizing predicted and actual property values in combination with the numeric scores to build a better intuition about the predictive performance.

Note that having a robust evaluation metric only partially solves the problem as ML models are susceptible to over-confident extrapolations. As we will show later, in imbalanced learning scenarios, ML models make overconfident extrapolations
which have higher probability of being wrong (e.g., predicting conductor to be an insulator with $99\%$ confidence). In other words, a model's own confidence score cannot be trusted. To overcome this problem, we use a set of labeled experimentally known compounds as side information to help determine a model's trustworthiness for a particular unseen test example. The trust score is defined as follows:

\begin{equation}
\label{trust_score}
T(X_i) = 1-\dfrac{d\left(X_i,\{X_j\}_{j\in c_i}\right)}{d\left(X_i,\{X_j\}_{j\in c_i}\right)+d\left(X_i,\{X_j\}_{j\notin c_i}\right)}.
\end{equation} 
The trust score $T$ takes into account the average Gower distance $d$ from the test sample $X_i$ to other samples in the same class $c_i$ vs. the average Gower distance to nearby samples in other classes. $T$ ranges from $0$ to $1$ where a higher $T$ value indicates a more trustworthy model.


\subsection{Example Applications}
In this section, we discuss two distinct applications for our reliable and explainable ML pipeline to demonstrate its versatility: predicting five physically distinct properties of crystalline compounds and identifying potentially stable solar cells. In both the cases, we use the same general framework, i.e., the same attributes and ML pipeline. Through these examples, we discuss all aspects of creating reliable and explainable ML models: building a reliable machine learning model from distributionally skewed training data, generating explanations to gain better understanding of the data/model, evaluating model accuracy and employing the model to predict new materials.

\subsubsection{Predicting Properties of Crystalline Compounds}
\label{application1}
Density functional theory (DFT) provides a means of predicting properties of chemical compounds, based on quantum mechanical modeling. However, the utility of DFT is limited by its computational complexity. An alternative approach is to use machine learning (ML) to train a surrogate model on a representative set of (input,output) pairs from prior DFT calculations. The surrogate then emulates DFT, producing approximate answers at dramatically lower computational cost (several orders of magnitude faster), enabling rapid screening of candidate materials. A potential drawback of this approach is that it requires many (potentially hundreds of thousands) DFT calculations in order to generate a suitable training set. Fortunately, several such training sets already exist.

\textbf{\textit{Data Set:}}
We follow the lead of Ward et al.\cite{ward2016general} and use OQMD for training purposes. OQMD contains the results of DFT calculations on approximately $300,000$ diverse compounds. Of these, we select only the lowest-energy compound for each composition. This yields a training set containing $228,573$ unique examples. We use the same set of $145$ attributes/features to represent each compound as Ward et al.\cite{ward2016general} Using these features, we consider the problem of developing reliable and explainable ML models to predict five physically distinct properties currently available through the OQMD: bandgap energy (eV), volume/atom (\r{A}$^3$/atom), energy/atom (eV/atom), thermodynamic stability (eV/atom) and formation energy (eV/atom)\cite{phy_prop}. 
Units for these properties are omitted in the rest of the paper for ease of notation.
A description of the $145$ attributes (inputs) and $5$ properties (outputs) are provided in the Supplementary Materials.

\textbf{\textit{Method:}}
We quantify the predictive performance of our approach using $5$-fold cross-validation. 
Following the procedure mentioned in Sec.~\ref{train_app}, we partition the property space for each property in $K=3$ classes. The decision boundary thresholds for class separation (with class distributions) are as follows: bandgap energy ($0.0,0.4$) with ($94\%, 5\%, 1\%$), volume/atom ($20.0,40.0$) with ($42\%, 55\%, 3\%$), energy/atom ($-8.0,-4.0$) with ($1\%, 63\%, 26\%$), stability ($0.0,1.5$) with ($8\%, 89\%, 3\%$) and formation energy ($0.0,1.0$) with ($40\%, 53\%, 7\%$).\footnote{We also tried different combinations of thresholds and trends in the obtained results were found to be consistent. In practice, these thresholds can be provided by domain experts depending on a specific application (as done in Sec.~\ref{sec:solarcell}).}. 
Sub-sampling ratios for sample compounds (for obtaining evenly distributed training set) were determined using cross-validation. We train Extreme Gradient Boosting (XGB) classifiers to do multi-class ($K=3$) classification using the softmax objective for each property. Next, we train Gradient Boosting Regressors (GBRs) for each property-class pair independently (and refer to them as marginal regressors). Using these marginal regressors, we create augmented feature vectors for correlation based predictions. Finally, we train another set of GBR regressors for each property-class pair on augmented data (and refer to them as joint regressors as they exploit correlation present among properties to improve the prediction performance).

\begin{table}[!t]
\caption{Results for conventional technique with overall prediction scores. Cross-validation gives an impression that conventional regressors have excellent regression performance (i.e., low MAE/MSE and high R$^2$ score). However, later we show that these metrics provide misleading inferences due to the presence of distributionally skewed data.}
\resizebox{1.0\textwidth}{!}{
    \begin{tabular}{| l | l | l | l | l | l |}
    \hline
    Metrics & Energy/atom & Volume/atom & Bandgap Energy & Formation Energy & Stability \\ \hline
    MAE & $0.1336$ & $0.5534$ & $0.0480$ & $0.0908$ & $0.0895$  \\ \hline
    MSE & $0.0349$ & $0.8727$ & $0.0542$ & $0.0198$ & $0.0281$ \\ \hline
    R$^2$ & $0.9915$ & $0.9861$ & $0.8865$ & $0.9709$ & $0.8657$ \\ \hline
    \end{tabular}
    }
	\label{table:conv_over}
\end{table}

\textbf{\textit{Results:}}
For the conventional scheme, we train $M$ independent GBR regressors to directly predict properties from the features corresponding to the compounds. In Table~\ref{table:conv_over}, we report different error metrics to quantify the regression performance using cross-validation. Note that these metrics report an accumulated/average error score on the test set (which comprises of compounds from all partitions of properties). These results are comparable to state of the art\cite{ward2016general} and suggest that conventional regressors have excellent regression performance (low MAE/MSE and high R$^2$ score). Relying on the inference made by this evaluation method, we may be tempted to use these regression models in practice for different applications (such as, screening or discovery of novel solar cells). However, next we show that these metrics provide misleading inferences in the presence of distributionally skewed data. In Table~\ref{table:conv_class}, we perform class specific evaluations (i.e., we partition the property space for each property in $K=3$ classes and use the test data belonging to each class separately). Surprisingly, Table~\ref{table:conv_class} shows that conventional regressors perform well only on a specific class (or range of property values) -- specifically, only those in the majority classes (i.e., majority of compounds fall in those property value ranges). The conventional regression method performs particularly poorly with minority classes for bandgap energy and stability prediction where the data distribution is highly skewed (see Fig.~\ref{imb_hist}). Unfortunately, the test data is also distributionally skewed and is not representative of the true data distribution. Thus, traditional methods for assessing and ensuring generalizability of ML models provide misleading conclusions (as shown in Table~\ref{table:conv_over}). On the other hand, class specific evaluations better characterize the predictive performance of ML models in the presence of distributionally skewed data.   

\begin{table}[!t]
\caption{Class-specific prediction score comparison. Class-specific cross-validation provides reliable inferences and shows the superiority of proposed scheme over conventional scheme.}
\vspace{-0.1in}
\subtable[Conventional technique. Class-specific cross-validation shows that the conventional technique performs poorly on minority classes. This important observation cannot be made from Table~\ref{table:conv_over}.]{ 
\resizebox{1.0\textwidth}{!}{
    \begin{tabular}{| l | l | l | l | l | l |}
    \hline
    Metrics & Energy/atom & Volume/atom & Bandgap Energy & Formation Energy & Stability \\
        Distribution & ($1\%, 63\%, 26\%$) & ($42\%, 55\%, 3\%$) & ($94\%, 5\%, 1\%$) & ($40\%, 53\%, 7\%$)& ($8\%, 89\%, 3\%$)\\\hline
    MAE & $(0.17,0.13,0.11)$ & $(0.46,0.58,1.3)$ & $(0.02,0.49,0.77)$ & $(0.10,0.08,0.14)$ & $(0.12,0.08,0.23)$  \\ \hline
    MSE & $(0.05,0.04,0.03)$ & $(0.52,0.83,6.3)$ & $(0.02,0.49,1.29)$ & $(0.02,0.01,0.07)$ & $(0.04,0.01,0.33)$ \\ \hline
    R$^2$ & $(0.94,0.97,0.95)$ & $(0.93,0.96,0.89)$ & $(1.0,0.55,-0.13)$ & $(0.97,0.80,0.47)$ & $(-0.86,0.86,-0.09)$ \\ \hline
    \end{tabular}}\label{table:conv_class}
}
\vspace{0.1in}
\subtable[Proposed technique without transfer learning. Simplicity (or explainability) due to smaller and simpler models results in performance loss. This is not surprising as there is a trade-off between simplicity/explainability and prediction performance.]{
\resizebox{1.0\textwidth}{!}{
    \begin{tabular}{| l | l | l | l | l | l |}
    \hline
    Metrics & Energy/atom & Volume/atom & Bandgap Energy & Formation Energy & Stability \\ \hline
    MAE & $(0.27,0.30,0.18)$ & $(0.64,0.93,2.3)$ & $(0.01,0.62,0.81)$ & $(0.16,0.13,0.20)$ & $(0.13,0.14,0.29)$  \\ \hline
    MSE & $(0.13,0.15,0.07)$ & $(0.89,1.84,10.8)$ & $(0.02,0.70,1.64)$ & $(0.05,0.03,0.10)$ & $(0.05,0.03,0.37)$ \\ \hline
    R$^2$ & $(0.84,0.91,0.74)$ & $(0.88,0.91,0.81)$ & $(1.0,0.37,-0.43)$ & $(0.92,0.54,0.22)$ & $(-1.5,0.68,-0.26)$ \\ \hline
    \end{tabular}}\label{table:proposed_wotl}
}
\vspace{0.1in}
\subtable[Proposed technique with transfer learning. Transfer learning step in our pipeline compensates for the performance loss due to simplicity of models and in fact outperforms conventional technique (especially on minority classes). We suspect that this gain may also be due to the fact that simpler models perform better in low-data regime (e.g, minority classes), as opposed to complex models which may over-fit (and require a large amount of data to perform well).]{
\resizebox{1.0\textwidth}{!}{ 
    \begin{tabular}{| l | l | l | l | l | l |} 
    \hline
    Metrics & Energy/atom & Volume/atom & Bandgap Energy & Formation Energy & Stability \\ \hline
    MAE & $(0.13,0.13,0.07)$ & $(0.39,0.54,1.2)$ & $(0.01,0.49,0.70)$ & $(0.08,0.08,0.2)$ & $(0.08,0.07,0.19)$  \\ \hline
    MSE & $(0.04,0.03,0.02)$ & $(0.43,0.80,6.7)$ & $(0.02,0.55,1.42)$ & $(0.02,0.01,0.07)$ & $(0.03,0.01,0.31)$ \\ \hline
    R$^2$ & $(0.95,0.98,0.93)$ & $(0.94,0.96,0.88)$ & $(1.0,0.51,-0.24)$ & $(0.97,0.82,0.46)$ & $(-0.36,0.87,-0.05)$ \\ \hline
    \end{tabular}}\label{table:proposed_wtl}
} 
\end{table}

In Table~\ref{table:proposed_wotl}, we show the effect of transforming a single complex regression model into ensemble of smaller and simpler models to gain a better understanding of sub-domains (Step $1$-$4$ in Sec.~\ref{train_app}). We notice that the performance of these transformed simpler models are worse compared to having a single complex model (as given in Table~\ref{table:conv_class}). This suggests that there is a trade-off between simplicity/explainability and accuracy.

Finally, Table~\ref{table:proposed_wtl} shows how this performance loss due to simplicity of models can be overcome using the transfer learning (or correlation based fusion) step in our pipeline. We observe that the proposed transfer learning technique can very well exploit correlations in the property space which results in a significant performance gain compared to conventional regression approach\footnote{Surprisingly, we did not observe any gain when using transfer learning with conventional technique. In fact, we observed that the models showed severe over-fitting behavior to the predicted properties.}. Note that this gain is achieved in spite of having simper and smaller models in our ML pipeline. This suggests that a user can achieve high accuracy without sacrificing explainability.  
We also observed that sub-sampling step in our pipeline had a positive impact on the regression performance of minority classes.

\begin{figure}[t]
	\centering
	\subfigure[]{
		\includegraphics[%
		width=0.45\textwidth,clip=true]{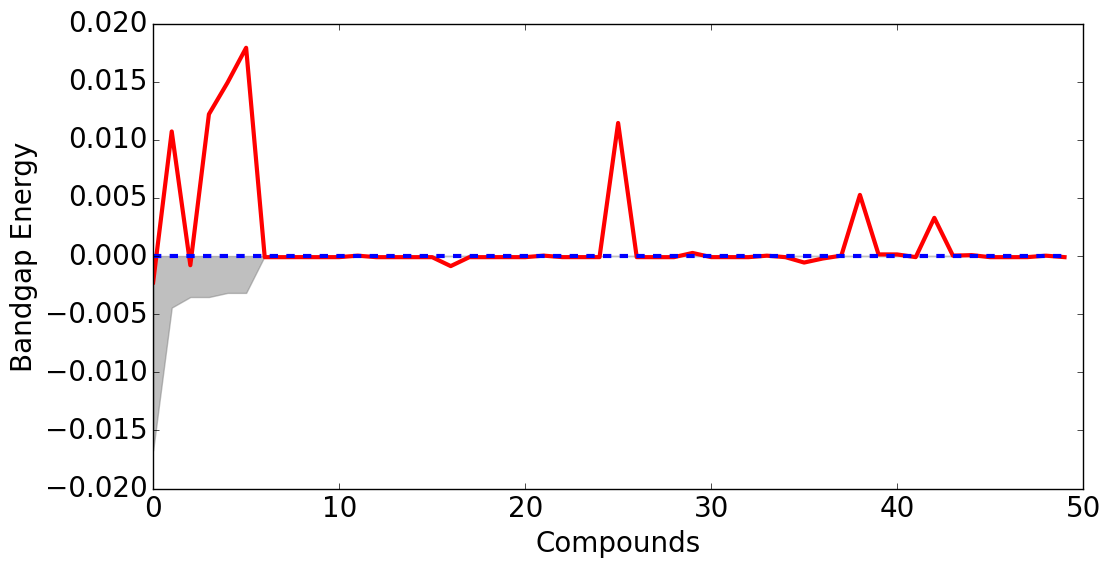}
		\label{bguq}}
	\hspace{0.1in}
	\subfigure[] {
		\includegraphics[%
		width=0.45\textwidth,clip=true]{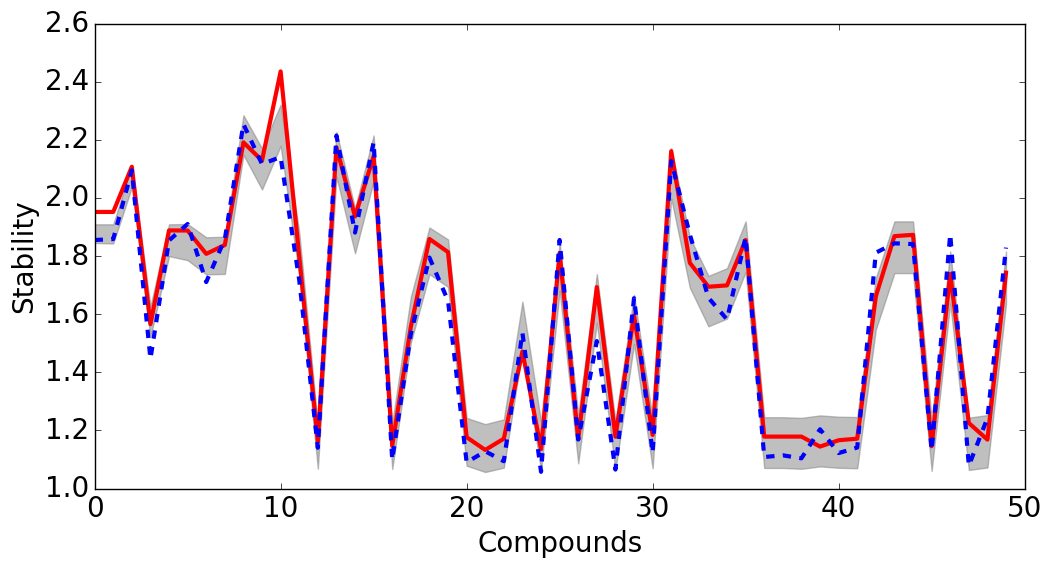}
  \label{stuq} }
	\caption{Uncertainty quantification of the regressor (ground truth is in blue, predictions are in red, and gray shaded area represents uncertainty). \subref{bguq} Bandgap energy, and \subref{stuq} Stability. In several cases, regressors perform poorly in regions with high uncertainty.}
	\label{uq}
	\vspace{-0.1in}
\end{figure}

Furthermore, our pipeline also quantifies uncertainties in its predictions providing a confidence score to the user. We show an illustration of the uncertainty quantification of bandgap energy and stability predictions on $50$ test samples in Figure~\ref{uq}. It can be seen that regressors perform poorly in regions with high uncertainty.

We would also like to point out that in cases where the data from a specific class is heavily under-represented, none of the model design strategies will improve the performance and generating new data may be the only possible solution (e.g., bandgap energy prediction for minority classes). In such cases, relying solely on cross-validation score or confidence score may not provide reliable inference (shown later). To overcome this challenge, explainable machine learning can be a potentially viable solution. 

\begin{figure}[t]
	\centering
	{
		\includegraphics[%
		width=0.85\textwidth,clip=true]{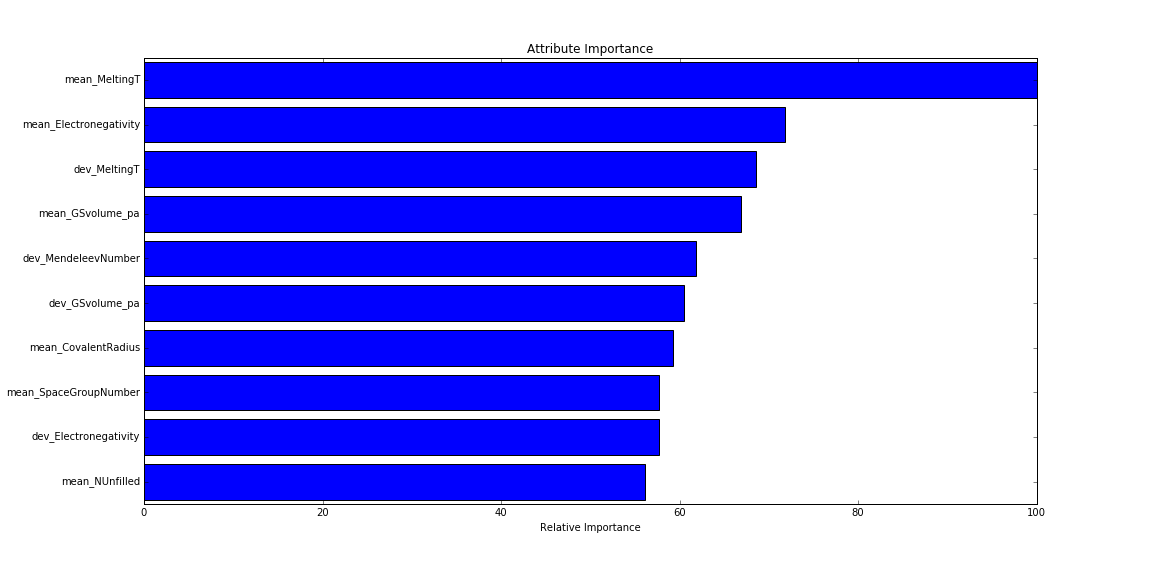}
		}
\vspace{-0.2in}
	\caption{Feature importance for $3$-class classification of bandgap energy. {The rationale generator favors attributes related to melting temperature, electro-negativity, and volume per atom for explaining bandgap-energy predictions. These attributes are all known to be highly correlated with the bandgap energy level of crystalline compounds.}}
	\label{fi}
	\vspace{-0.1in}
\end{figure}

Next, we show the output of rationale generator in our pipeline. Specifically, we provide $1)$ model-level explanations, as well as, $2)$ decision-level explanations for each sub-class of materials.  
For model-level explanations, our pipeline provides feature importance for both classification and regression steps.
Feature importance provides a score that indicates how useful (or valuable) each feature was in the construction of the model. The more an attribute is used to make key decisions with (classification/regression) model, the higher its relative importance. This importance is calculated explicitly for each attribute in the data set, allowing attributes to be ranked and compared to each other.
In Fig.~\ref{fi}, we show the feature importance for our $3$-class classifier for bandgap energy. It shows the attributes which help in discriminating among $3$-classes on compounds (insulators, semi-conductors, and conductors) based on their bandgap energy values.
Note that the rationale generator picked attributes related to the melting temperature, electro-negativity and volume per atom of constituent elements to be the most important features in determining the bandgap energy level of the compounds. This is reasonable as all these attributes are known to be highly correlated with the bandgap energy level of crystalline compounds. {For example, melting temperature of constituent elements is positively correlated with inter-atomic forces (and in turn inter-atomic distances).
Increased inter-atomic spacing decreases the potential seen by the electrons in the material, which in turn reduces the bandgap energy. Therefore, band structure changes as function of inter-atomic forces which is correlated with melting temperature. Similarly, in multi-element material system, as the electro-negativity difference between different atoms increases, so does the energy difference between bonding and anti-bonding orbitals. Therefore, the bandgap energy increases as the electro-negativities of constituent elements increase. Thus, the bandgap energy has a strong correlation with electro-negativity of constituent elements. Finally, mean volume per atom of constituent elements is also correlated with the inter-atomic distance in a material system. As explained above, inter-atomic distance is negatively correlated with the bandgap energy, and so does the mean volume per atom of constituent elements.
}
Similar feature importance results for class-specific predictors can also be obtained (see Supplementary Material).

\begin{table}[!t]
\caption{Bandgap energy prediction and uncertainty quantification. Model's own confidence score alone may not be very reliable as it makes wrong and over-confident predictions on minority classes (i.e., classes $1$ and $2$). On the other hand, a higher (or lower) trust score consistently imply higher (or lower) probability that the classifier (or regressor) is correct.}
\resizebox{0.8\textwidth}{!}{
    \begin{tabular}{| l | l | l | l | l |}
    \hline 
    Test Compound & Ground Truth & Prediction & Confidence Score & Trust Score \\ 
     & (Class, Bandgap) & (Class, Bandgap) &  &  \\    \hline
    $\text{Ge}_1\text{Na}_1\text{O}_3$ & $(2,4.63)$ & $(0,0.0)$ & $0.999$ & $0.43$  \\ \hline
    $\text{F}_6\text{Na}_2\text{Nb}_1$ & $ (1,0.1)$ & $(0,0.0)$ & $0.998$ & $0.42$ \\ \hline 
    $\text{C}_2\text{Mg}_1$ & $(2,2.67)$ & $(0,0.0)$ & $0.998$ & $0.27$ \\ \hline
	$\text{Rh}_1$ & $(0,0.0)$ & $(0,0.0)$ & $0.999$ & $0.80$ \\ \hline
    \end{tabular}
    }
	\label{trust}
\end{table}
In Table~\ref{trust}, we show $4$ test compounds with ground truths (class, bandgap energy value), predictions (class, bandgap energy value), and corresponding confidence scores. It can be seen that both classifier and regressor make wrong and over-confident predictions on minority classes (i.e., classes $1$ and $2$). 
In other words, a higher confidence score from the model for minority class does not necessarily imply higher probability that the classifier (or regressor) is correct. For compounds in minority classes, ML model may simply not be the best judge of its own trustworthiness. On the other hand, the proposed trust score (as given in \eqref{trust_score}) consistently outperforms classifier's/regressor's own confidence score. A higher/lower trust score from the model imply higher/lower probability that the classifier (or regressor) is correct. Furthermore, as our trust score is computed using distances from experimentally known compounds from Inorganic Crystal Structure Database (ICSD)\cite{icsd}, it also provides some confidence on compounds amenability to be synthesized. 


\subsubsection{Novel Stable Solar Cell Prediction}
\label{sec:solarcell}
To show how our ML pipeline can be used
for discovering new materials, we simulate a search for
stable compounds with bandgap energy within a desired range. 
To evaluate the ability of our approach to locate compounds that are stable and have bandgap energies within the target range, we setup an experiment where a model was fit on the training data set and, then, was tasked with selecting which $30$ compounds in the test data were most likely to be stable and have a bandgap energy in the desired range for solar cells: $0.9 - 1.7$ eV. 

\textbf{\textit{Data Set:}}
Same as before, for the training data, we selected a subset of $228,573$ compounds from OQMD that represents the lowest-energy compounds at each unique composition. We use same $145$ attributes as before. Using these attributes/features, we consider the problem of developing reliable and explainable ML models to predict two physically distinct properties of stable solar cells: bandgap energy, and stability. 
Note that this experiment is more challenging and practical as compared to Ward et al.\cite{ward2016general} where the training data set was considered to be compounds that were reported to be possible to be made experimentally in the ICSD (a total of $25,085$ entries) so that only bandgap energy, and not stability, needed to be considered. 
We choose test data set from Meredig et al.\cite{ms3} to be as-yet-undiscovered ternary compounds ($4,500$ entries). These compounds are not yet in the OQMD.

\begin{table}[!t]
\caption{Compositions of materials predicted using proposed ML pipeline to be stable candidates for solar cell applications with experimentally known prototypes and their distances from predicted candidates.}
\resizebox{1.0\textwidth}{!}{
    \begin{tabular}{| l | l | l | l | l | l |}
    \hline 
    Compounds &  Bandgap Energy & Stability & (Prototype $1$, Distance) & (Prototype $2$, Distance) & Trust Score \\ \hline
    $\text{Cr}_1\text{Se}_4\text{Cs}_4$ & $1.20$ & $-0.10$ & $(\text{Cs}_2\text{Mn}_1\text{Se}_2,0.03)$ & $(\text{Rb}_2\text{Mn}_1\text{Se}_2,0.06)$ & $0.52$  \\ \hline
    $\text{S}_1\text{Sb}_3\text{Cs}_3$ & $1.38$ & $-0.09$ & $(\text{Cs}_3\text{Ge}_1\text{Te}_3,0.06)$ & $(\text{Cs}_2\text{Si}_1\text{As}_2,0.07)$ & $0.49$ \\ \hline 
    $\text{V}_1\text{Se}_4\text{Cs}_3$ & $1.47$ & $-0.21$ & $(\text{Cs}_2\text{Zr}_1\text{Se}_3,0.02)$ & $(\text{Cs}_2\text{Nb}_1\text{Ag}_1\text{Se}_4,0.04)$ & $0.53$ \\ \hline
	$\text{C}_1\text{O}_2\text{Th}_2$ & $1.28$ & $-0.16$ & $(\text{Th}_1\text{S}_1\text{O}_1,0.07)$ & $(\text{Th}_2\text{S}_1\text{N}_2,0.07)$ & $0.58$ \\ \hline
		$\text{Se}_3\text{Pm}_{1.33}\text{Pt}_1$ & $1.28$ & $-0.11$ & $(\text{Sm}_1\text{Cu}_1\text{Se}_2,0.06)$ & $(\text{Ho}_1\text{Ag}_1\text{Se}_2,0.07)$ & $0.50$ \\ \hline
	$\text{O}_5\text{Na}_9\text{Ag}_1$ & $1.41$ & $-0.007$ & $(\text{Na}_{14}\text{Cd}_2\text{O}_9,0.02)$ & $(\text{Na}_3\text{Ag}_1\text{O}_2,0.03)$ & $0.59$ \\ \hline

    \end{tabular}
    }
	\label{solar_new}
\end{table}

\textbf{\textit{Method:}}
Following the procedure mentioned in Sec.~\ref{train_app}, we partition the property space for each property in $K=3$ classes. The decision boundary thresholds for class separation are as follows: bandgap energy ($0.9,1.7$), and stability ($0.0,1.5$). Similar to Sec.~\ref{application1}, we use Extreme Gradient Boosting (XGB) classifiers (with default parameters) to do multiclass ($K=3$) classification and Gradient Boosting Regressors (GBRs) to do marginal and joint regression. We use models' own confidence and trust score to rank the potentially stable solar cells. 


\textbf{\textit{Results:}}
We used the proposed ML pipeline to search for new stable compounds (i.e., those not yet in the OQMD). Specifically, we use trained models to predict bandgap energy and stability of compositions that were suggested by Meredig et al.\cite{ms3} to be as-yet-undiscovered ternary compounds. We found that out of these $4500$ compounds, $221$ compounds are likely to be stable and have favorable bandgap energies to be solar cells. A subset with the trust score are shown in Table~\ref{solar_new}. Similar experimentally known prototypes (as shown in Table~\ref{solar_new}) can also serve as an initial guess on the $3$-d crystal structure of the predicted compounds. 
These recommendations appear reasonable as four of the six suggested compounds (Cs$_4$CrSe$_4$, Cs$_3$Sb$_3$S, Cs$_3$VSe$_4$, Na$_9$AgO$_5$) can be classified as I-III-VI semiconductors, which are semiconductors that contain an alkali metal, a transition metal, and a chalcogen; I-III-VI semiconductors are a known promising class of photovoltaic materials as many have direct bandgap energies of $\sim 1.5$ eV, making them well-matched to the solar spectrum. The best known I-III-VI photovoltaic is copper-indium-gallium-selenide (CIGS), which has solar cell power conversion efficiencies on par with silicon’s. The other two identified compounds – Th$_2$CO$_2$ and Pm$_{1.33}$PtSe$_3$ – are unique in that they contain actinide and lanthanide elements. However, from a practical perspective, the scarcity and radioactivity of these elements may make it challenging to explore them experimentally.
A detailed list of potentially stable solar cell compounds is provided in the Supplementary Material. 

{\section{Some Open Issues}
There are still some issues yet to be resolved for a successful application of ML in material science. 
First, in cases where the data from a specific class is heavily under-represented, none of the model design strategies will improve the performance and generating new data may be the only possible solution. Solving this problem will require answering the following  question: How many training samples are sufficient to learn a reliable model and where to sample if they are inadequate?
Second, predictive models built based on chemical attributes make recommendations (e.g., potential solar cells) in the form of chemical attributes. However, verifying these recommendations using DFT (or experiments) has its own challenges (e.g., identifying appropriate crystal structure (or synthesis recipes)). A potentially viable solution is to bias the recommendation process towards compounds with favourable synthesis conditions.  
Finally, explainable ML methods based on feature importance still require a material scientist to make sense of model/decision explanations using domain knowledge which may suffer from the human bias. Solving these problems will require making significant advances on current explainable ML techniques. Interactive ML and casual inference techniques can further help in resolving some of these issues.
}
\section{Conclusions}
This paper considered the problem of learning reliable and explainable machine learning models from underrepresented and distributionally skewed materials science data. We identified common pitfalls of existing ML techniques while learning from imbalanced data. We show how applying ML techniques without careful consideration of its assumptions and limitations can lead to both quantitatively and qualitatively
incorrect predictive models. To overcome the limitations of existing ML techniques, we proposed a general-purpose explainable and reliable ML framework for learning from imbalanced material data. We also proposed 
a new evaluation metric and a trust score to better quantify confidence in the predictions. The rationale generator component in our pipeline provides useful model-level and decision-level explanations to establish trust in the ML model and its predictions. Finally, we demonstrated the applicability of our technique on predicting five physically distinct properties of crystalline compounds, and identifying
potentially stable solar cells.

\section{Materials and Methods}
All machine learning models were created using the Scikit-learn\cite{scikit} and XGBoost\cite{xgb} machine learning libraries. The Materials Agnostic Platform for Informatics and Exploration (Magpie)\cite{ward2016general} was used to compute the attributes. Scikit-learn, XGBoost and Magpie are available under open-source licenses. The software, training data sets and input files used in this work are provided in the Supplementary Information associated with this manuscript.

\section{Acknowledgements}
Authors would like to thank Dr. Joel Varley and Dr. Mike Surh for valuable feedbacks, suggestions and discussions in preparation of this manuscript.

This work was performed under the auspices of the U.S. Department of Energy by Lawrence Livermore National Laboratory under Contract DE-AC52-07NA27344 and was supported by the LLNL-LDRD Program under Project No. 16-ERD-019 and 19-SI-001. (LLNL-JRNL-764864)
\section{Contributions}
B.K. and T.Y.H conceived the project, B.K performed the experiments, B.K. and B.G. analyzed the results. All authors discussed the results and contributed to the writing of the manuscript.
\section{Competing Interests}
The authors declare no conflict of interest

\section{Data Availability}
All data generated or analysed during this study are included in this published article (and its supplementary information files).

\bibliographystyle{IEEEtran}
\bibliography{references}

\newpage
\section*{Supplementary Material}
\subsection{Attributes and Properties}
The first step of our pipeline is to compute attributes (or chemical descriptors) based on the composition of materials. These attributes should be descriptive enough to enable a ML algorithm to construct general rules that can possibly ``learn"
chemistry. Building on existing
strategies\cite{ward2016general}, we use a set of $145$ attributes/features to represent each compound. These attributes are comprised of: stoichiometric properties, elemental statistics, electronic structure properties attributes, ionic compound attributes. A detailed procedure to compute these attributes can be found in The Materials Agnostic Platform for Informatics and Exploration
(Magpie)\cite{ward2016general}. 
Using these features, we consider the problem of developing reliable and explainable ML models to predict five physically distinct properties currently available through the OQMD: bandgap energy (eV), volume/atom (\r{A}$^3$/atom), energy/atom (eV/atom), thermodynamic stability (eV/atom) and formation energy (eV/atom). Formation energy is total Energy/atom minus some correction factors (i.e., the material with the lowest formation energy at each composition also has the lowest energy per atom). Stability has to do with whether a particular material is thermodynamically stable or not. Compounds with a negative stability are stable and those with a positive stability are unstable. More information on the properties are provided by Emery et al.\cite{property, phy_prop}.
\vspace{-0.1in}
\subsection{Feature Importance for Class-specific Regression}
Feature importance results for class-specific predictors can also be obtained.

\begin{figure}[h]
 	\centering
 	\subfigure[]{
 		\includegraphics[%
 		width=0.8\textwidth,clip=true]{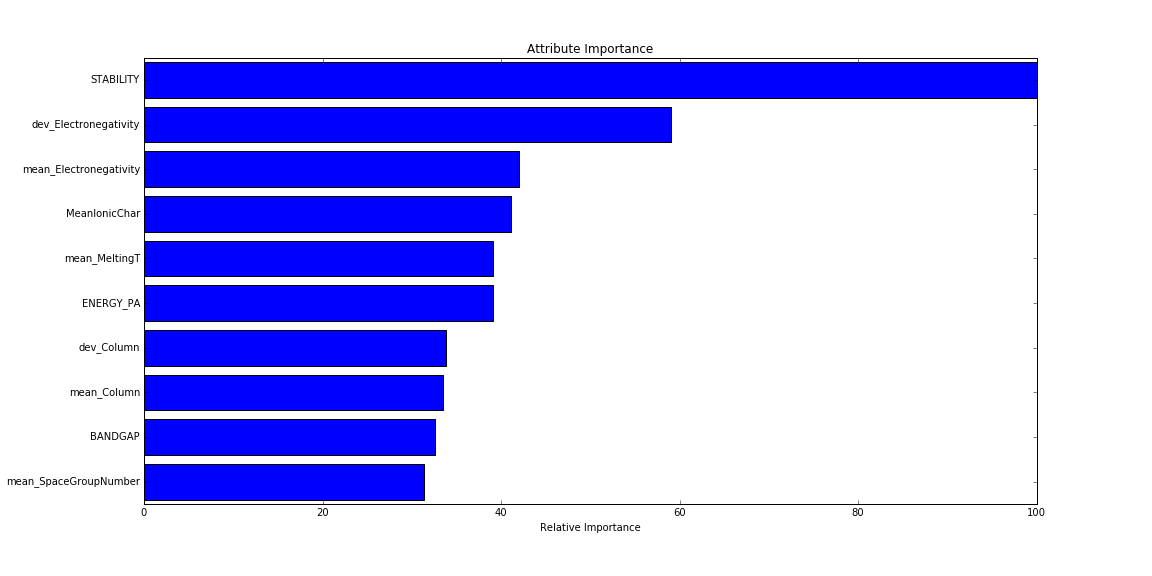}
 		\label{fe0fi}}
 	\hspace{-0.2in}
 		\subfigure[]{
 		\includegraphics[%
 		width=0.8\textwidth,clip=true]{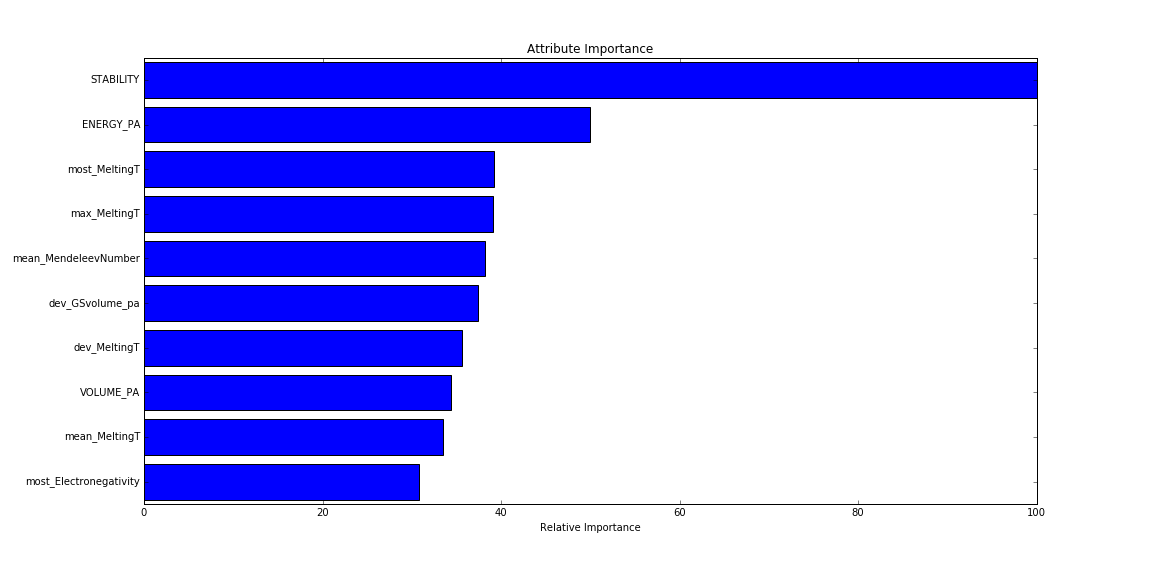}
 		\label{fe1fi}}
 	\hspace{-0.2in}
 	\subfigure[] {
 		\includegraphics[%
 		width=0.8\textwidth,clip=true]{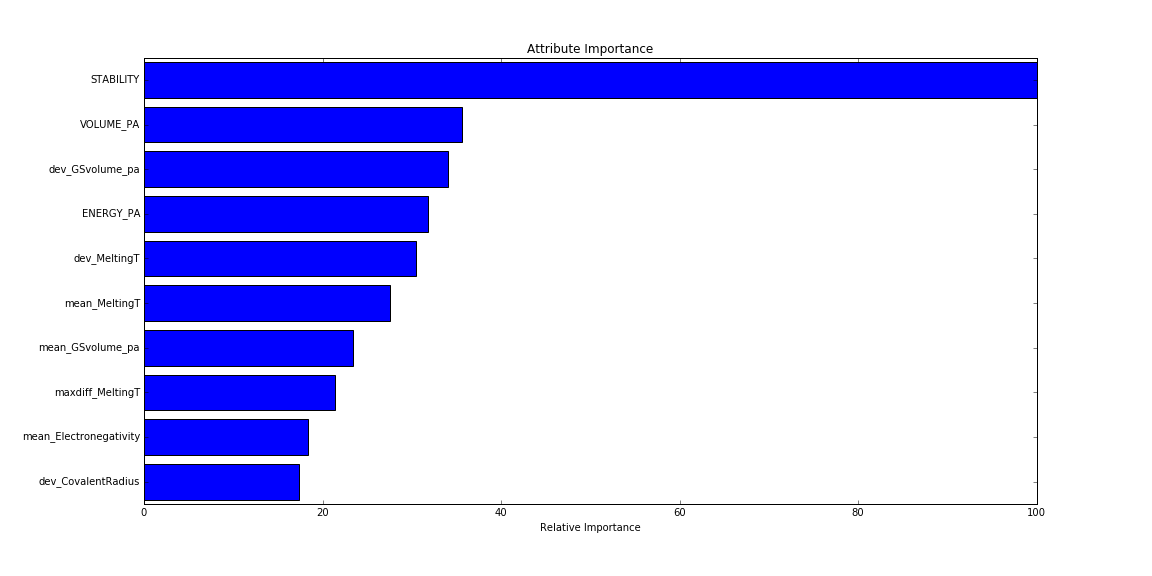}
 		\label{fe2fi} }
 	\caption{Feature importance for class specific formation energy prediction regressors. \subref{fe0fi} class $0$, \subref{fe1fi} class $1$, and \subref{fe2fi} class $2$,.}
 	\label{fecfi}
 	\vspace{-0.1in}
 \end{figure}

In Fig.~\ref{fecfi}, we show feature importance for formation energy prediction regressors for all $3$ classes. 
{{For all three classes, the thermodynamic stability is found to be the most important attribute in predicting formation energy. From thermodynamic point of view, this makes sense as the stability is negatively correlated with the formation energy.}}
More results are provided in the Supplementary Information associated with this manuscript.
\vspace{-0.1in}

\subsection{Stable Solar Cell Compounds}
A detailed list of potentially stable solar cell (with corresponding property
predictions and explanations) is provided in the Supplementary Information associated with this manuscript.
\vspace{-0.1in}

\subsection{Other}
The software, training data sets and input files used in this work are provided in the Supplementary Information associated with this manuscript.

\end{document}